\documentclass{iopart}
\usepackage[latin1]{inputenc}
\usepackage{iopams}
\usepackage{color}
\usepackage{graphicx}
\usepackage{hyperref}
\usepackage{ulem}
\usepackage{cite}

\definecolor{battleshipgrey}{rgb}{0.52, 0.52, 0.51}
\definecolor{cadet}{rgb}{0.33, 0.41, 0.47}
\definecolor{charcoal}{rgb}{0.21, 0.27, 0.31}

\hypersetup{
	colorlinks   = true, 
	urlcolor     = battleshipgrey, 
	linkcolor    = cadet, 
	citecolor   = charcoal 
}

\newcommand{\ip}{I_\mathrm{p}}
\newcommand{\up}{U_\mathrm{p}}
\renewcommand\Re{\mathrm{Re}}

\begin{document}

\title{Coulomb-free and Coulomb-distorted recolliding quantum orbits in photoelectron holography}

\author{A. S. Maxwell\footnote{andrew.maxwell.14@ucl.ac.uk}
	and C. Figueira de Morisson Faria\footnote{c.faria@ucl.ac.uk} }
\date{\today}

\begin{abstract}
	We perform a detailed analysis of the different types of orbits in the Coulomb Quantum Orbit Strong-field Approximation (CQSFA), ranging from direct to those undergoing hard collisions. We show that some of them exhibit clear counterparts in the standard formulations of the strong-field approximation for direct and rescattered above-threshold ionization, and show that the standard orbit classification commonly used in Coulomb-corrected models is over-simplified. We identify several types of rescattered orbits, such as those responsible for the low-energy structures reported in the literature, and determine the momentum regions in which they occur. We also find formerly overlooked interference patterns caused by backscattered, Coulomb-corrected orbits and assess their effect on photoelectron angular distributions. These orbits improves the agreement of photoelectron angular distributions computed with the CQSFA with the outcome of ab-initio methods for high-energy phtotoelectrons perpendicular to the field-polarization axis. 
\end{abstract}

\maketitle



\section{Introduction}

Laser-induced recollisions have played a vital role in strong-field phenomena such as above-threshold ionization (ATI) for over two decades \cite{Corkum1993}. The quintessential example of a recollision-induced effect is the ATI plateau, which, for a monochromatic, linearly polarized field, or a long enough pulse,  may extend to a photoelectron energy of up to $10U_p$, where $U_p=I/(4\omega^2)$ is the ponderomotive energy, $I$ denotes the driving-field intensity  and $\omega$ gives the laser frequency. This structure has been first identified in the mid 1990s \cite{Paulus1994} (for reviews see, e.g., \cite{Becker2002Rev,Agostini2012Rev}), and consists of ATI peaks with comparable intensities.  It results from a hard elastic collision of an electron, which is backscattered by its parent ion \cite{Paulus1994b,Lewenstein1995}. Up to recently, it was accepted knowledge that scattering in ATI was only important for intermediate and high photoelectron energy ranges. For energies up to around $2U_p$, the ATI plateau is obfuscated by the contribution from the so-called direct electrons, which result from strong-field ionization in the absence of recollision with the core. 

 Recently, however, this has been called into question in \cite{Moller2014,Becker2015,Milos2016}, in which it has been shown, within the framework of the Strong-Field Approximation (SFA), that rescattering is also important for much lower photoelectron energy. This has been attributed to the large scattering cross section that is specific to the Coulomb potential for solutions of the ATI transition amplitude that had been previously overlooked. Among these, the importance of forward-scattered trajectories has been highlighted. In particular, low-energy rescattering events lead to a wide range of structures in photoelectron velocity maps that have been previously identified in experiments, such as a cusp-like, low-energy structure \cite{Blaga2009,Quan2009}, a fork-like structure \cite{Moller2014} and a pronounced V-shaped structure \cite{Wolter2014}. 

The quantum interference of rescattered with direct electrons, or of trajectories associated with different types of rescattering, has been paramount for the development of time-resolved photoelectron holography. The key idea behind it is that there is a probe and reference signal, which are associated with different types of trajectories, whose interference leads to a wide range of patterns. Well-known examples are the spider-like structure that results from the interference of different types of forward-scattered trajectories\cite{Huismans2011,Huismans2012}, the fishbone structure caused by the interference of direct and the backscattered electron
wavepackets \cite{Huismans2011,Bian2011}, and the near-threshold fan-shaped structure caused by the interference of direct and forward-deflected trajectories \cite{Rudenko2004b,Maharjan2006}.
The very concept of ``direct" and ``rescattered" electrons has been defined either using classical models, for which the residual binding potentials are neglected during the electron propagation, or the strong-field approximation, for which the continuum is approximated by field-dressed plane waves and the core by a single point at the origin. This implies that such models allow for the existence of either hard collisions, or no collisions at all. 

In a realistic situation, however, the residual binding potential does influence the electron propagation in the continuum. Thus, the difference between direct and rescattered electron trajectories is blurred, and there may be direct trajectories, deflected trajectories, soft and hard collisions. A key question is how to determine whether a specific electron orbit should be viewed as ``direct" or ``rescattered", using the terminology implicit in the SFA, and what types of rescattering can be identified. Only in the case of low-energy structures have soft collisions been categorized in terms of phase-space criteria \cite{Kastner2012,Kastner2012PRL,Kelvich2016}.

In recent publications, we have explored Coulomb effects in photoelectron holography using the newly developed Coulomb Quantum-Orbit Strong-Field Approximation (CQSFA) \cite{Lai2015,Lai2017,Maxwell2017,Maxwell2018}. We have shown that the fan-shaped pattern that forms near the threshold, may be viewed as a holographic structure stemming from the interference between direct and forward deflected electron trajectories.  Furthermore, we have successfully reproduced the spider-like structure caused by the interference  of two different types of forward deflected trajectories, and have identified a myriad of other patterns that have been overlooked in the literature. 
Analytical estimates have also been used to single out direct, deflected and soft-scattered trajectories. They have shown that scattering plays a vital role in the spider-like structure, and that, in the absence of hard collisions, the acceleration caused by the residual Coulomb potential is the key contributor to extending the ATI signal beyond the direct-ATI cutoff of $2U_p$. Previous work has also shown the existence of a specific type of orbit, which, as the photoelectron energy increased, moved from a field-induced deflection to a hard scattering process \cite{Maxwell2018}. 

The main objective of this work is to
investigate the role of scattering for the Coulomb-corrected orbits encountered in the CQSFA, in comparison with the standard SFA. We will focus on the low- and intermediate photoelectron energy ranges investigated in \cite{Becker2015}, due to the rich structures encountered in this regime. We will also address how different types of trajectories in the CQSFA relate to their Coulomb-free, SFA counterparts. Thereby, we will use the same classification as in \cite{Yan2010a} which singles out four different types of orbits, with emphasis on how the initial and final momentum components relate to each other. This article is organized as follows. In Sec.~\ref{sec:backgrd}, we provide the necessary theoretical background. Subsequently, in Sec.~\ref{sec:orbs}, we perform a detailed analysis of the CQSFA orbits as compared to the standard direct and rescattered orbits that are present in the SFA, and refine the standard orbit classification employed in Coulomb-corrected models. In Sec.~\ref{sec:cutoff}, we investigate the overall shapes present in single-orbit probability distributions and establish the momentum regions occupied for the CQSFA and rescattered SFA. In Sec.~\ref{sec:interf}, we study previously overlooked interference features associated with backscattered trajectories and perform comparisons with ab-initio methods. Finally, in Sec.~\ref{sec:concl} we state our conclusions. Unless otherwise stated, we use atomic units throughout. 

\section{Background}
\label{sec:backgrd}
\subsection{Strong-field approximation for direct and rescattered electrons}
Below we will briefly recall the SFA transition amplitudes for direct and rescattered ATI, which will be compared with the outcome of the CQSFA. For more details we refer to \cite{Lohr1997}  and the review \cite{Becker2002Rev}.
The direct SFA transition amplitude considers a transition from an initial bound state $|\Psi_{0}(t')\rangle=\exp[iI_pt']|\Psi_{0}\rangle$ to a continuum state $|\mathbf{p}+\mathbf{A}(t)\rangle=U^{(V)}(t,t')|\mathbf{p}+\mathbf{A}(t')\rangle$, which is approximated by a field-dressed plane wave, without further interaction with the core. Thereby, $U^{(V)}(t,t')$ is the Volkov time-evolution operator, which is related to a field-dressed free particle, $I_p$ gives the system's ionization potential, V is the binding potential and $\mathbf{A}(t)$ the vector potential associated to the external laser field. Explicitly, 
\begin{equation}
M_{d}(\mathbf{p})=-i \int_{-\infty}^{\infty}dt' \langle \mathbf{p}+\mathbf{A}(t')|V|\Psi_{0}\rangle e^{i S_{d}(\mathbf{p},t')},
\label{eq:dirSFA}
\end{equation}
where
\begin{equation}  \label{Action}
S_d(\mathbf{p},t')= -\frac{1}{2} \int^{\infty}_{t'}[\mathbf{p}+\mathbf{A}(\tau)]^2 d\tau + I_pt'
\end{equation}
is the action describing the above-mentioned process. 

If, on the other hand, one incorporates up to a single act of rescattering, the SFA transition amplitude reads
\begin{equation}
M_{r}(\mathbf{p})=- \int^{\infty}_{-\infty} dt \int^{t}_{-\infty} dt'\int d^3k\exp[iS_{r}(\mathbf{p},\mathbf{k},t,t')]V_{\mathbf{k}0}V_{\mathbf{pk}},
\label{eq:rescSFA}
\end{equation}
where the action is given by
\begin{equation}
S_{r}(\mathbf{p},\mathbf{k},t,t')=-\frac{1}{2} \int^{\infty}_{t}[\mathbf{p}+\mathbf{A}(\tau)]^2 d\tau -\frac{1}{2} \int^{t}_{t'}[\mathbf{k}+\mathbf{A}(\tau)]^2 d\tau + I_pt'
\label{eq:Sresc}
\end{equation}
and all the influence of the core is incorporated in the ionization prefactor
\begin{equation}
V_{\mathbf{k}0}=\left\langle \mathbf{k}+\mathbf{A}(t)\right|V \left| \Psi_0 \right \rangle
\end{equation}
and in the rescattering prefactor 
\begin{equation}
V_{\mathbf{pk}}=\left\langle \mathbf{p}+\mathbf{A}(t)\right|V \left| \mathbf{k}+\mathbf{A}(t) \right \rangle.
\end{equation}
Eq.~(\ref{eq:rescSFA}) gives the transition amplitude associated with a process in which an electron, initially in a bound state $|\Psi_{0}(t')\rangle$, is freed an instant $t'$, propagates in the continuum with an intermediate momentum $\mathbf{k}$ for $t'<\tau<t$ and recollides with its parent ion at a later time $t$. Upon recollision, it then acquires the final momentum $\mathbf{p}$.
Eqs.~(\ref{eq:dirSFA}) and (\ref{eq:rescSFA}) are solved using the steepest descent method. This method requires that the actions $S_d(\mathbf{p},t')$ and $S_{r}(\mathbf{p},\mathbf{k},t,t')$ be stationary. The saddle-point equation obtained from the direct action $S_d(\mathbf{p},t')$ reads
\begin{equation}
[\mathbf{p}+\mathbf{A}(t')]^2 = - 2I_pt',
\label{eq:tunnelsfa}
\end{equation}
which expresses the kinetic energy conservation at the time of ionization. One should note that Eq.~(\ref{eq:tunnelsfa}) has no real solutions, which reflects the fact that tunnel ionization has no classical counterpart. A formally identical equation is obtained by imposing the condition $\partial S_{r}(\mathbf{p},\mathbf{k},t,t')/\partial t'=0$, with the final momentum $\mathbf{p}$ being replaced by the intermediate momentum $\mathbf{k}$. This gives
\begin{equation}
\label{eq:tunnelsfaresc}
[\mathbf{k}+\mathbf{A}(t')]^2 = - 2I_pt'.
\end{equation}
The condition $\partial S_{r}(\mathbf{p},\mathbf{k},t,t')/\partial t=0$ yields the conservation of energy
\begin{equation}
[\mathbf{p}+\mathbf{A}(t)]^2=[\mathbf{k}+\mathbf{A}(t)]^2
\label{eq:saddresc}
\end{equation}
upon recollision, and $\partial S_{r}(\mathbf{p},\mathbf{k},t,t')/\partial \mathbf{k}=\mathbf{0}$ leads to the constraint
\begin{equation}
\mathbf{k}=-\frac{1}{t-t'}\int_{t}^{t'}\mathbf{A}(\tau)d\tau
\label{eq:saddk}
\end{equation}
upon the intermediate momentum $\mathbf{k}$, such that it returns to the site of its release, i.e., the origin. Eqs.~(\ref{eq:saddresc}) and (\ref{eq:saddk}) in fact imply that the electron suffers a hard collision as it returns to the origin. These are the same equations employed in \cite{Becker2015}. We compute the direct transition amplitude (\ref{eq:dirSFA}) using the standard saddle point approximation, and the rescattering transition amplitude (\ref{eq:rescSFA}) employing the specific uniform approximation discussed in \cite{Faria2002}. Throughout, we will employ the acronyms DATI and HATI for direct and high-order, rescattered ATI, respectively. 
\subsection{The Coulomb quantum orbit strong-field approximation}
\label{sec:cqsfa}
We will now provide a brief outline of the Coulomb Quantum Orbit Strong-Field Approximation (CQSFA). For more details we refer to our previous publications \cite{Lai2015,Lai2017,Maxwell2017}. In order to account for the Coulomb potential when the electron is the continuum, we employ a path integral method on the transition amplitude \begin{equation}\label{eq:Mpp}
M(\mathbf{p}_f)\hspace*{-0.1cm}=\hspace*{-0.1cm}-i \lim_{t\rightarrow \infty}\hspace*{-0.15cm}
\int_{-\infty }^{t }\hspace*{-0.2cm}d t'\hspace*{-0.2cm}
\int d \mathbf{\tilde{p}}_0 \left\langle  \mathbf{\tilde{p}}_f(t)
|U(t,t') |\mathbf{\tilde{p}}_0\right \rangle \left \langle
\mathbf{\tilde{p}}_0 | H_I(t')| \Psi
_0(t')\right\rangle \, 
\end{equation}
from an initial bound state $\left | \Psi_0(t')\right\rangle $ to a final continuum state
 $|\mathbf{\tilde{p}}_f(t)\rangle=|\psi_{\mathbf{p}}(t)
\rangle$. The variables  $\mathbf{\tilde{p}}_0=\mathbf{p}_0+\mathbf{A}(t')$ and $\mathbf{\tilde{p}}_f(t)=\mathbf{p}_f+\mathbf{A}(t)$ 
give the initial and final velocity of the electron at the times $t'$ and $t$, respectively. The time evolution operator $U(t,t')$ is related to the full Hamiltonian $H(t)=H_a+H_I(t)$, where
\begin{equation}
H_a=\frac{\hat{\mathbf{p}}^{2}}{2}+V(\hat{\mathbf{r}})
\end{equation}
gives the field-free one-electron atomic Hamiltonian and $\hat{\mathbf{r}}$ and $\hat{\mathbf{p}}$ denote the position and momentum operators, respectively, and $H_I(t)$ gives the interaction with the external field. The binding potential is taken to be of Coulomb type, i.e., 
\begin{equation}
V(\hat{\mathbf{r}})=-\frac{1}{\sqrt{\hat{\mathbf{r}}\cdot
		\hat{\mathbf{r}}}},\label{eq:potential}
\end{equation} 
 and  the  interaction Hamiltonian is chosen to be in the length gauge, so that
\begin{equation}
H_I(t)=-\hat{\mathbf{r}}\cdot \mathbf{E}(t),
\end{equation}
 where $\mathbf{E}(t)=-d\mathbf{A}(t)/dt $ is the external laser field. Eq.~(\ref{eq:Mpp}) incorporates the full continuum dynamics for the system. It is however inaccurate for transitions involving bound states, such as excitation and relaxation. 
 
This leads to the expression
 \begin{eqnarray}\label{eq:pathintegral}
 M(\mathbf{p}_f)&=&-i\lim_{t\rightarrow \infty
 }\int_{-\infty}^{t}dt' \int d\mathbf{\tilde{p}}_0
 \int_{\mathbf{\tilde{p}}_0}^{\mathbf{\tilde{p}}_f(t)} \mathcal {D}'
 \mathbf{\tilde{p}}  \int
 \frac{\mathcal {D}\mathbf{r}}{(2\pi)^3}  \nonumber \\
 && \times  e^{i S(\mathbf{\tilde{p}},\mathbf{r},t,t')}
 \langle
 \mathbf{\tilde{p}}_0 | H_I(t')| \psi _0  \rangle \, ,
 \end{eqnarray}
 where $\mathcal{D}'\mathbf{p}$ and $\mathcal{D}\mathbf{r}$ are the integration measures for the path integrals, and the prime indicates a restriction. The action reads as
 \begin{equation}\label{eq:stilde}
 S(\mathbf{\tilde{p}},\mathbf{r},t,t')=I_pt'-\int^{t}_{t'}[
 \dot{\mathbf{p}}(\tau)\cdot \mathbf{r}(\tau)
 +H(\mathbf{r}(\tau),\mathbf{p}(\tau),\tau)]d\tau,
 \end{equation}
 and
 \begin{equation}
 H(\mathbf{r}(\tau),\mathbf{p}(\tau),\tau)=\frac{1}{2}\left[\mathbf{p}(\tau)+\mathbf{A}(\tau)\right]^2
 +V(\mathbf{r}(\tau)),
 \label{eq:Hamiltonianpath}
 \end{equation}
 where the intermediate momentum and position have been parametrized in terms of the time $\tau$.
Physically, Eq.~(\ref{eq:pathintegral}) represents a sum  over all possible paths available to the electron in position and momentum between its start and end points. We also seek solutions for the variables $t'$, $\mathbf{p}$ and $\mathbf{r}$ such that the action is stationary. This leads to the equation
\begin{equation}
\frac{\left[\mathbf{p}(t')+\mathbf{A}(t')\right]^2}{2}+V(\mathbf{r}(t'))=-I_p,
\label{eq:tunncc}
\end{equation}
which is the Coulomb-corrected counterpart of Eq.~(\ref{eq:tunnelsfa}), 
and to the equations
\begin{equation}
\mathbf{\dot{p}}=-\nabla_rV(\mathbf{r}(\tau))\label{eq:p-spe}
\end{equation}
and
\begin{equation}
\mathbf{\dot{r}}= \mathbf{p}+\mathbf{A}(\tau),\label{eq:q-spe}
\end{equation}
which are the classical equations of motion of the electron in the continuum. 

 The transition amplitude (\ref{eq:pathintegral}) is computed using a two-pronged contour, whose first and second parts are parallel to the imaginary and real-time axis, respectively. The first part of the contour, from $t'=t'_r+it'_i$ to $t'_r$, describes tunnel ionization, and the second part of the contour, from $t'_r$ to $t$, describe continuum propagation. This specific contour has been widely used in the implementation of Coulomb-corrected approaches\cite{Popruzhenko2008,Yan2012,Torlina2012,Torlina2013}. Inside the barrier, we neglect the influence of the Coulomb potential on the electron momentum, which is kept as $\mathbf{p}_0$.

Thus, the binding potential is neglected in Eq.~(\ref{eq:tunncc}), which then becomes formally identical to its SFA counterpart (\ref{eq:tunnelsfa}). The action in the first arm of the contour reads
\begin{equation}
S^{\mathrm{tun}}(\mathbf{\tilde{p}},\mathbf{r},t'_r,t')=I_p(it'_i)-\frac{1}{2}\int_{t'}^{t'_r}\left[ \mathbf{p}_0+\mathbf{A} (\tau)\right]^2d\tau  -\int_{t'}^{t'_r}V(\mathbf{r}_0(\tau))d\tau, \label{eq:stunn}
\end{equation}where $\mathbf{r}_0$ is defined by 
\begin{equation}
\mathbf{r}_0(\tau)=\int_{t'}^{\tau}(\mathbf{p}_0+\mathbf{A}(\tau'))d\tau',
\label{eq:tunneltrajectory}
\end{equation} 
which is widely known as ``the tunnel trajectory". The action $S^{\mathrm{tun}}(\mathbf{\tilde{p}},\mathbf{r},t'_r,t')$ inside the barrier is calculated from the origin until the tunnel exit
\begin{equation}\label{exit}
z_0=\Re[r_{0z}(t'_r)].
\end{equation}

The action in the second arm of the contour is given by 
\begin{equation}
S^{\mathrm{prop}}(\mathbf{\tilde{p}},\mathbf{r},t,t'_r)=I_p(t_r)-\frac{1}{2}\int_{t'_r}^{t}\left[ \mathbf{p}(\tau)+\mathbf{A} (\tau)\right]^2d\tau - 2\int_{t'_r}^{t}V(\mathbf{r}(\tau))d\tau, \label{eq:sprop2}
\end{equation}
where the factor 2 in front of the Coulomb integral stems from the fact that $\mathbf{r} \cdot \dot{\mathbf{p}}=V(r)$ \cite{Maxwell2017,Shvetsov-Shilovski2016b}. The total action is then given by \begin{equation}
S(\mathbf{\tilde{p}},\mathbf{r},t,t')=S^{\mathrm{tun}}(\mathbf{\tilde{p}},\mathbf{r},t'_r,t')+S^{\mathrm{prop}}(\mathbf{\tilde{p}},\mathbf{r},t,t_r').
\end{equation}
Within the saddle-point approximation, the transition amplitude is written as
	\begin{equation}
	\label{eq:MpPathSaddle}
	M(\mathbf{p}_f)\propto-i \lim_{t\rightarrow \infty } \sum_{s}\bigg\{\det \bigg[  \frac{\partial\mathbf{p}_s(t)}{\partial \mathbf{r}_s(t_s)} \bigg] \bigg\}^{-1/2} \hspace*{-0.6cm}
	\mathcal{C}(t_s) e^{i
		S(\mathbf{\tilde{p}}_s,\textbf{r}_s,t,t_s)} ,
	\end{equation}where $t_s$, $\mathbf{p}_s$ and $\mathbf{r}_s$ are determined by the above-stated saddle point equations and $\mathcal{C}(t_s)$ is given by
\begin{equation}
\label{eq:Prefactor}
\mathcal{C}(t_s)=\sqrt{\frac{2 \pi i}{\partial^{2}	S(\mathbf{\tilde{p}}_s,\textbf{r}_s,t,t_s) / \partial t^{2}_{s}}}\langle \mathbf{p}+\mathbf{A}(t_s)|H_I(t_s)|\Psi_{0}\rangle.
\end{equation}
  In practice, we use  $\partial
\mathbf{p}_s(t)/\partial \mathbf{p}_s(t_s)$ instead of the stability factor stated in Eq.~(\ref{eq:MpPathSaddle}), which may be obtained employing a Legendre transformation. As long as the electron starts from the origin, the action will not be modified by this choice. We normalize  Eq.~(\ref{eq:MpPathSaddle}) so that the SFA transition amplitude is obtained in the limit of vanishing binding potential, and take the electron to be initially in a $1s$ state \cite{Lai2015}.  

\subsection{Electron Orbits}

The formulation described in the previous section allows for  all kinds of recollision, and four main types of orbits. These orbits have first been classified in \cite{Yan2010a} according to the tunnel exit and their initial and final momentum components. Let us call the momentum components parallel and perpendicular to the driving-field polarization $p_{j\parallel}$, $p_{j\perp}$, respectively, with $j=0,f$. 
\begin{table} [h!]
	\begin{center}
		\begin{tabular}{ c c c } 
			\hline
			\hline 
			Orbit & $z_0p_{f\parallel}$ & $p_{f\perp}p_{0\perp}$ \\ 
			\hline
			\hline
			1 & + & + \\ 
			2 & - & + \\ 
			3 & - & - \\ 
			4 & + & - \\ 
			\hline
		\end{tabular}
	\end{center}
	\caption{Summary of the main types of orbits identified for Coulomb-corrected strong-field approaches. The + and - signs on each cell indicate a positive or negative product, respectively.}\label{tab:orbits}
\end{table}

For type 1 orbits, the electron leaves from a tunnel exit on the same side as the detector, i.e., $z_0p_{f\parallel}>0$, and the transverse momentum component does not change sign, so that  $p_{f\perp}p_{0\perp}>0$. These are the direct orbits, for which rescattering may not occur and along which the electron will be decelerated by the residual binding potential. In contrast, for type 2 and 3 orbits, the electron is freed on the opposite side of the target, i.e., $z_0p_{f\parallel}<0$, with the difference that the transverse momentum changes sign for obit 3, but remains in the same direction for orbit 2. These orbits are known as forward-scattered trajectories, and this scattering may in principle range from a light deflection to a hard collision. In particular, we have verified that Orbit 3 is considerably accelerated by the Coulomb potential, and it is not possible to reproduce many features associated with the spider-like structures if we do not include at least some type of recollision. This has been shown in our previous publication \cite{Maxwell2018}, in which analytical models have been constructed for disentangling different features in the CQSFA. Finally, for orbit 4, the electron leaves in the direction of the detector, but its transverse momentum changes direction. This means that it will undergo some kind of backscattering. Since, however, the transition amplitude associated with this orbit was small in the parameter range of interest, it has not been investigated in our previous work \cite{Lai2015,Maxwell2017,Maxwell2018}. For clarity, in Table \ref{tab:orbits}, we provide a summary of how the different types of orbits behave.

\section{Orbit Properties}
\label{sec:orbs}
\begin{figure}
	\includegraphics[width=\textwidth]{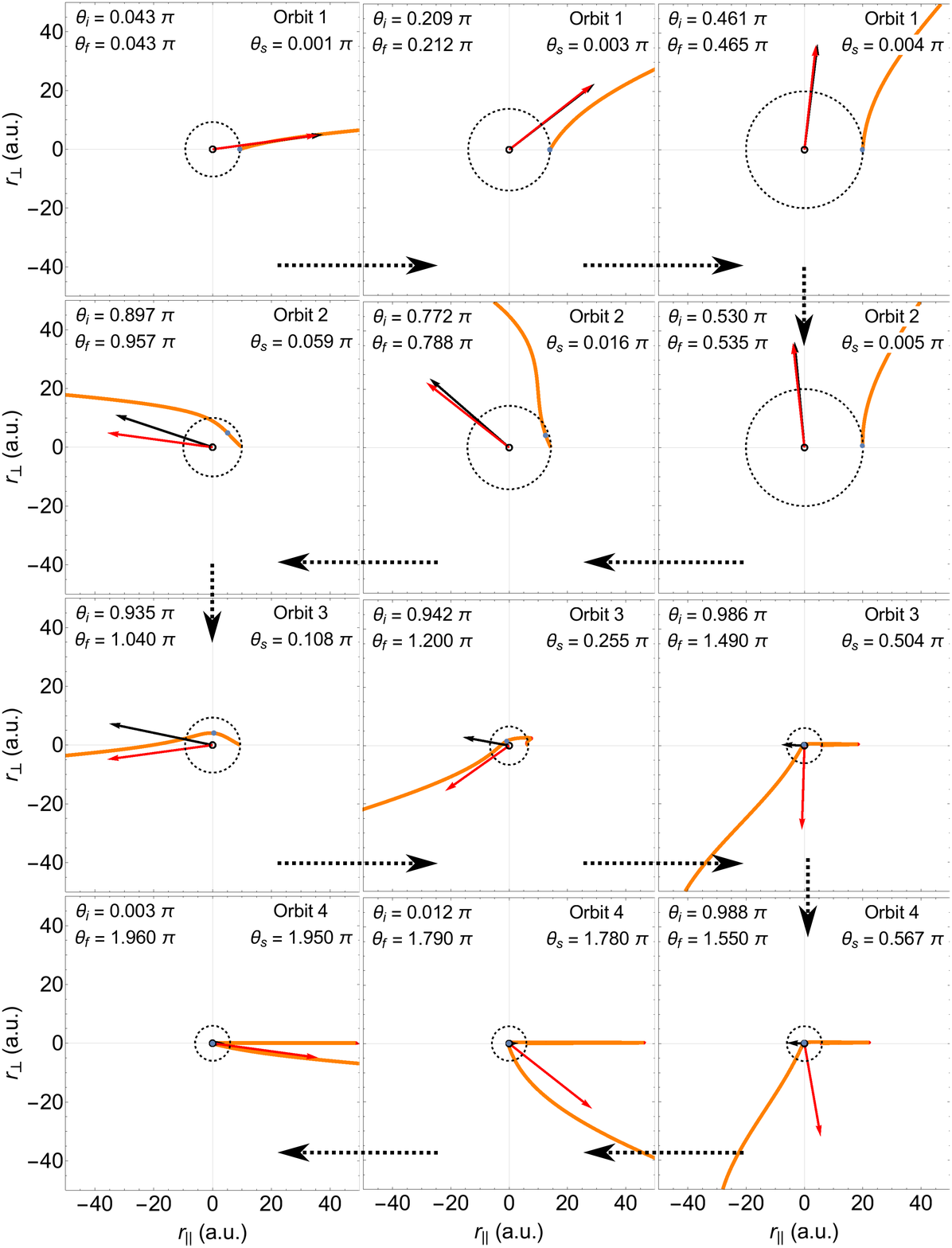}
	\caption{Electron trajectories calculated using equations (\ref{eq:p-spe}) and (\ref{eq:q-spe}) for fixed energy, $E = 1.3$ a.u. or $E=3 \up$, and values of the angle  $\theta_f$ associated with the final momenta in the range $0<\theta_f<2\pi$. 
	Each orange line represents an electron trajectory after tunnelling, while the dotted arrows point in the direction of increasing $\theta_f$. The initial and final momentum vectors are marked on the figure by black and red arrows, respectively. The angles related to the initial and final momentum are given by $\theta_i$ and $\theta_f$ (top left), while their difference is given by $\theta_s$ (top right). The orbit type, defined in Sec.~\ref{sec:backgrd} and summarised in Table \ref{tab:orbits}, is marked in the top right corner of each panel. As guides, circles whose radii are the Bohr radius and the tunnel exit are marked in solid black and in dashed black, respectively.  We consider a field of wavelength $\lambda=800 $ nm, intensity $I_0=2\times 10^{14}$ W/cm$^2$ and a model atom with ionisation potential $\ip=0.5$ a.u. This gives a ponderomotive energy of $\up=0.439$ a.u.}
	\label{fig:AllOrbs}
\end{figure}

The orbit classification introduced in the previous section is however too coarse, as we will argue next. This statement is enabled by an improved method of solving the saddle-point equations (\ref{eq:tunnelsfa}), (\ref{eq:p-spe}) and (\ref{eq:q-spe}). This is exemplified in Fig.~\ref{fig:AllOrbs}, by keeping the initial conditions fixed for a particular orbit and varying the final angle $\theta_f$ at which the final momentum $\mathbf{p}_f$ is detected from 0 to $2 \pi$. 
The initial conditions are chosen so that the tunnel exit and the perpendicular initial momentum are positive, ($z_0>0$, $p_{0\perp}>0$). Thus, following the classification used in \cite{Yan2010a}, the orbits will be categorised as type 1, 2, 3 and 4 for $0<\theta_f<\pi/2$, $\pi/2<\theta_f< \pi$, $\pi<\theta_f<3 \pi/2$ and $3\pi/2<\theta_f<2\pi$, respectively.

As $\theta_f$ moves from one quadrant to the other, the electron trajectories change smoothly from 1 to 4. Hence, the orbits are degenerate precisely at the boundaries between quadrants, i.e. on the axes. These boundaries are crossed when the thick dashed arrows in the figure become vertical. In fact, a resemblance can be seen very clearly for orbits 1 and 2, as the boundary $\theta_f=\pi/2$ is crossed (see upper two far right panels in Fig.~\ref{fig:AllOrbs}). This also holds for orbits 2 and 3, as the boundary $\theta_f=\pi$ is crossed (see left panels for which $\theta_f=0.957\pi$ and $\theta_f=1.040\pi$), and for orbits 3 and 4, as the boundary $\theta=3/2\pi$ is crossed (two lower far right panels).  Orbits 1 and 4 cannot be degenerate at $p_{f||}=0$, $\theta_f=2\pi, 0$, despite the electron's initial and final momentum being in the same direction for the two orbits. This is because in orbit 4 the electron undergoes a full $2\pi$ rotation around the ion, making the orbits qualitatively different.

The degeneracy between the four types of orbits at the boundaries can be exploited when solving the saddle point equations (\ref{eq:tunnelsfa}), (\ref{eq:p-spe}) and (\ref{eq:q-spe}). Starting from orbit 1 up to orbit 4, one computes each orbit up to the boundary, and uses it to provide initial conditions for the subsequent orbit type. For example, orbit 1 can be used to provide initial conditions for orbit 2 in the neighbourhood of $\theta_f=\pi/2$. Next, orbit 2 may be used near $\theta_f=\pi$ for solving orbit 3 and so on. This property is more useful with increasing orbit number, as the CQSFA orbits behave less like the DATI orbits in the standard SFA. In fact, this approach is essential for obtaining convergent solutions for orbit 4.

Given that the CQSFA spans these different orbit types, we would like to explore the idea that the CQSFA provides orbits that lie qualitatively between the two extremes of DATI (no collisions) and  HATI (hard collisions). In Fig.~\ref{fig:AllOrbs} one can see that for low final angle $\theta_f$ the electron trajectories behave qualitatively similarly to DATI trajectories. The electron does not revisit the ion and the initial and final momentum are almost the same. For high $\theta_f$ the electron trajectories resemble HATI orbits. The electron revisits the core, passing very close to the origin and undergoing what looks like a ``hard'' collision. There is almost no perpendicular momentum during the collision, in agreement with the saddle-point equation (\ref{eq:saddk}) that gives the electron's intermediate momentum within the SFA. 
The orbits in-between these angles are less well defined and may be strongly deflected or undergo soft collisions such as those seen in \cite{Kastner2012PRL,Kastner2012,Pisanty2016}. As such, the CQSFA can be seen to blur the distinction between direct and re-scattered ATI and softly colliding orbits.

\begin{figure}
	\includegraphics[width=\textwidth]{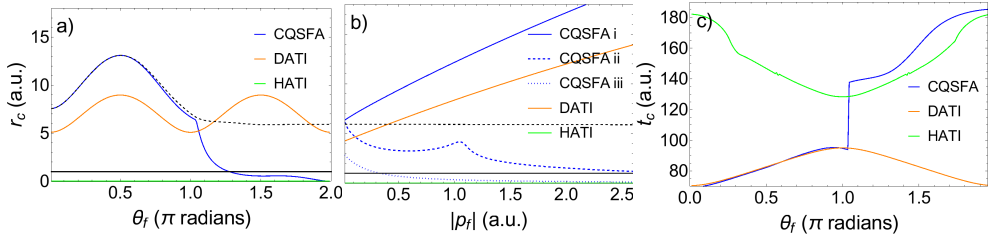}
	\caption{Distances $r_c$ and time $t_c$ of closest approach are plotted for the CQSFA, DATI and HATI.
		 In panel a), the distance of closest approach $r_c$ is plotted for a fixed energy $E=0.26$ a.u., and increasing angle $\theta_f$. The tunnel exit for the CQSFA is plotted in a black dashed line and the Bohr radius is marked with a solid black line. In panel b), $r_c$ is plotted for three fixed angles for the CQSFA. The indices i, ii and iii correspond to the angles, $0.25\pi$, $1.10\pi$ and $1.75\pi$,  respectively. The tunnel exit for the CQSFA in case iii is plotted with a black dashed line. For the  DATI and HATI, $r_c$ is plotted for a fixed angle of $\theta_f=1.75 \pi$. In panel c), the time of closest approach $t_c$ is plotted for a fixed energy of $E=0.26$ a.u. for the CQSFA, DATI and HATI. The same field and atomic parameters have been used as in Fig.~\ref{fig:AllOrbs}. For HATI we compute $\theta_f$ as given in \cite{Becker2015}.} 
	\label{fig:ClosestApproach}
\end{figure}

In addition, the CQSFA will behave more like the SFA for high energy orbits as there will be less interaction time with the core. We have in fact verified, in a previous publication \cite{Maxwell2017}, that the direct SFA is the high-energy limit of the CQSFA for orbits 1 and 2. 
 The angle $\theta_f$ will determine whether the CQSFA orbit will tend towards its DATI (low angles) or HATI (high angles) counterpart.  In the CQSFA, there can never be any truly ``hard'' collision (except in the limiting case $\theta_f=2\pi$) as the electron trajectory will always miss the origin by some amount. 
 In order to make a comparison with the DATI and HATI limits we have identified three important parameters, which determine the kind of collision we are dealing with:
 \begin{enumerate}
 	\item The Bohr radius, whose perimeter is  marked on Fig.~\ref{fig:AllOrbs} as a solid black circle, and which is indicated in Figs.~\ref{fig:ClosestApproach} a) and b) as a solid line.
 	\item  A circle whose radius is the tunnel exit, displayed in Fig.~\ref{fig:AllOrbs} in dashed black. This radius is also plotted as the black dashed lines in Figs.~\ref{fig:ClosestApproach} a) and b).
 	\item The electron's distance $r_c$ of closest approach after tunnelling, indicated in Fig.~\ref{fig:AllOrbs} by a blue spot.
 \end{enumerate}
  If an electron trajectory goes within the region determined by the Bohr radius, we assume it will undergo a ``hard'' collision as then it will interact as strongly with the core as a bound electron. Furthermore, orbits that do not enter region (ii) can be called direct as outside this perimeter the laser field dominates strongly over the potential. Finally, if the electron's trajectory closest point as defined in (iii) is between regions (i) and (ii), one may view it as softly recolliding.   
 Using these radii as a guide, one can see that orbit 1 may be always classified as a direct electron trajectory. In contrast, an electron along orbit 2 goes from direct to softly rescattered with increasing $\theta_f$. Orbit 3 will change from softly recolliding to a hard collision. Finally, for the parameter range of the figure, orbit 4 always corresponds to a hard collision.

\begin{figure}
	\includegraphics[width=\textwidth]{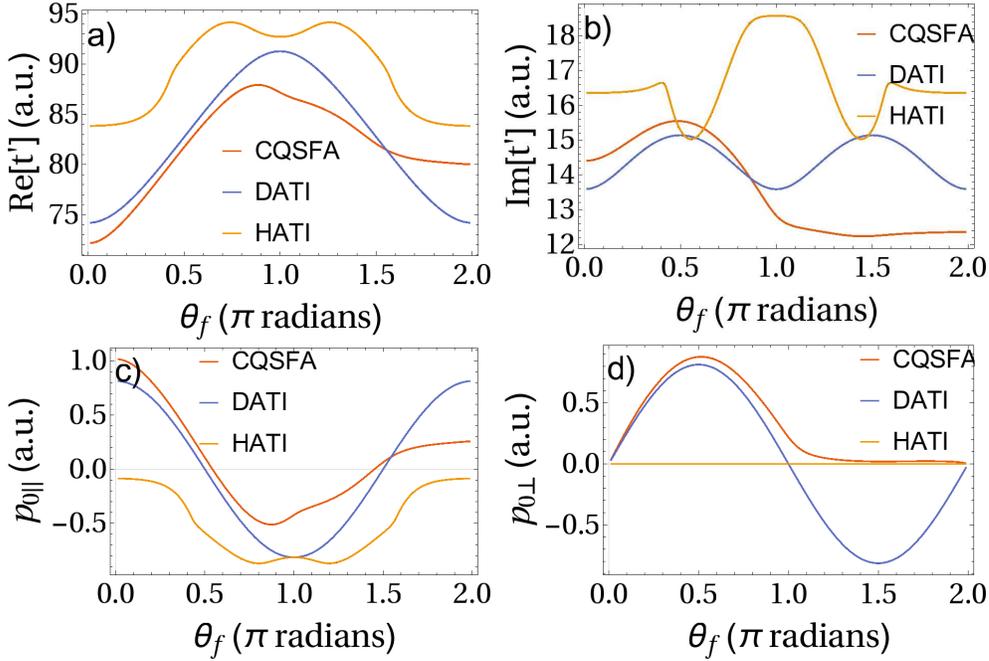}
	\caption{Ionisation times and initial momenta for the CQSFA, DATI and HATI, for  a fixed photoelectron energy $E=0.055$ a.u. and the same field and atomic parameters as in Fig.~\ref{fig:AllOrbs}. In panels a) and b), the real and imaginary parts of the time of ionisation are plotted. In panels c) and d),  we display the components of the initial momentum parallel and perpendicular to the laser-field polarisation.}
	\label{fig:Initials}
\end{figure}

In addition to the above-stated parameters, one may also use the time $t_c$ associated with $r_c$, the time of ionisation and the initial momentum $\mathbf{p}_0$ to compare the CQSFA orbits with the DATI and HATI models. These are presented in Figs.~\ref{fig:ClosestApproach} and \ref{fig:Initials}.
In Fig.~\ref{fig:ClosestApproach} a), we plot $r_c$ for fixed photoelectron energy as a function of the final angle $\theta_f$. For the CQSFA at low angles, $r_c$ (blue line) is the same as the tunnel exit (black dashed line) and the behaviour will mimic DATI (orange line). However, at $\theta_f= \pi/2$, when the electron trajectory becomes a type 2 orbit, the distance of closest approach moves away from the tunnel exit as the electron goes into the softly re-colliding region (see Fig.~\ref{fig:AllOrbs}).
 At $\theta_f\approx 1.1 \pi$, 
  there is a change in behaviour, marked by a discontinuous derivative in $r_c$. As $\theta_f$ increases further, the distance of closest approach falls rapidly, until it reaches the Bohr radius, where it levels off. This is not related to the change in orbit type from 2 to 3 at $\theta_f=\pi$. Instead, it is due to the nature of the orbit changing from deflection to hard collision. For $\pi<\theta_f<1.1\pi$, the electron moves entirely through the soft recolliding region defined by (ii)
  and the trajectory has a similar initial and final momentum. In contrast, for $\theta_f> 1.3 \pi$, the trajectory is initially moved outside of the softly recolliding region by the laser field. Subsequently, the field drives back the electron into the hard collision region. In this latter case, 
 the initial and final momentum are quite different as the electron gains  energy in the collision and is scattered through a larger angle. This kind of trajectory closely resembles HATI orbits. This transition can also be seen in the third row of Fig.~\ref{fig:AllOrbs}. A similar behaviour is displayed for the time $t_c$ of closest approach in Fig.~\ref{fig:ClosestApproach} c). For $t_c$ at low angles the CQSFA very closely follows the DATI ionisation time, with a discontinuity for $\theta_f\approx 1.1 \pi$. The time of closest approach jumps up at this point as the electron trajectory is now first taken away from the core before revisiting it. Hence, the closest approach occurs later in the orbit.  After the transition,  $t_c$ follows the HATI recollision time. 
     
     The CQSFA distance of closest approach $r_c$ is also plotted for three fixed angles over increasing photoelectron momenta in Fig.~\ref{fig:ClosestApproach} b). For $\theta_f=0.25\pi$, the CQSFA follows the DATI curve and stays on its tunnel exit. 
      For $\theta_f=1.1 \pi$, in the region where orbit 3 undergoes a transition, the CQSFA trajectory remains a softly colliding orbit for all energies staying just above the Bohr radius. For  $\theta_f=1.75 \pi$, the CQSFA orbit behaves like a HATI orbit and quickly falls below the Bohr radius with increasing energy.
      Nonetheless, one should note that, for very low energy, this is a softly-colliding orbit. In this region it behaves similarly to the softly colliding orbits discussed in \cite{Kastner2012,Kastner2012PRL} that are responsible for the low-energy structure (LES).

In Fig.~\ref{fig:Initials} we plot the initial time and momentum components for the CQSFA, DATI and HATI (upper and lower panels, respectively). 
In panel a), one can see that the real parts of the ionization times $t'$ are different for the  three approaches. This is due to the effect of the Coulomb potential, that favours shorter times for the CQSFA. However, as before, for low $\theta_f$ the CQSFA follows the DATI line, but as $\theta_f$ increases the behaviour tends towards the HATI curve.
Similarly, in panel b) $\mathrm{Im}[t']$ is different for the three approaches. This is because the Coulomb potential alters the tunnelling probability and hence shifts the imaginary component of the tunnelling time. Nonetheless, the qualitative behaviour of the CQSFA outcome also mirrors DATI for low angles and HATI for high angles. The same angular behaviour is present for the parallel and perpendicular initial momentum components $p_{0\parallel}$ and $p_{0\perp}$. In the perpendicular momentum case, the CQSFA momentum even tends asymptotically to its DATI and HATI counterparts as $\theta_f \rightarrow 0$ and $\theta_f \rightarrow 2 \pi$, respectively. It is remarkable that this behaviour is already present for low photoelectron energy, such as that employed in Fig.~\ref{fig:Initials}. In  all cases, we also see that the reflection symmetry about $\theta_f=\pi$ that exists in the Coulomb-free cases breaks for the CQSFA. This is expected as the dynamics of the system is no longer determined by the laser field alone. 

Other types of orbits that have been made accessible by our new solving method and which are important in the low-energy regime include those with multiple passes and/or trajectories whose dynamics are mainly determined by the Coulomb potential. 
In our previous work \cite{Maxwell2017} we stated that the soft-recolliding forward-scattered trajectories that form the inner spider \cite{Hickstein2012} and low energy structure \cite{Pisanty2016,Kastner2012,Kastner2012PRL,Kelvich2016} bear similarity to some type 3 orbits. Furthermore, we can make the same statement for softly colliding back-scattered trajectories and some type 4 orbits. 

\begin{figure}
	\includegraphics[width=\textwidth]{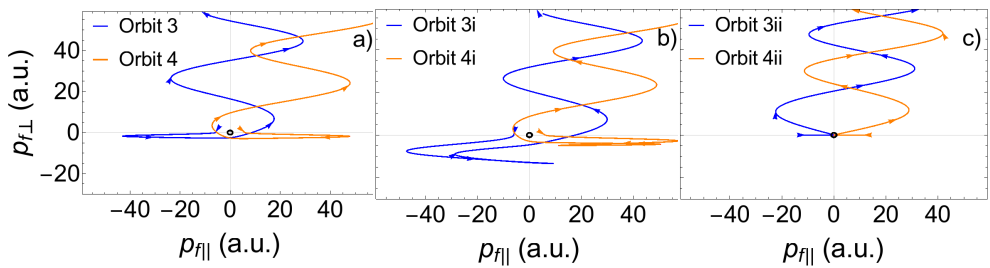}
	\caption{Different subtypes of orbit 3 and 4 that can occur for a final momentum $\mathbf{p}_f=(0.086,0.22)$ a.u., computed using the CQSFA. Panel a), b) and c) show the standard trajectories, the multi-pass orbits (denoted i) and the directly recolliding orbits (denoted ii), respectively.  The Bohr radius is marked by a black circle. The same field and atomic parameters have been used as in Fig.~\ref{fig:AllOrbs}.}
	\label{fig:OrbTypes}
\end{figure}

Fig.~\ref{fig:OrbTypes} in fact illustrates that classifying orbits into type 3 or type 4 is an over-simplification. Therein, we show three examples of orbits with very distinct dynamics, the same final momenta and which fall into the same classification according to the criteria in Table \ref{tab:orbits}. Apart from the standard case for these orbits, in panel a), there are type 3 and 4 orbits that are driven past the core many times (multi-pass) before softly scattering (see panel b)),  and also directly recolliding trajectories that hard-scatter off the core before the laser field has time to change sign (see panel c)).
As these orbits will lead to the same final momentum, they will potentially  interfere. However, combining these orbits would require a careful analysis of the orbits and potentially a new asymptotic expansion, which is beyond the scope of this work. The low energy (multi-pass) orbits are similar to longer HATI orbits that undergo a few passes before colliding with the core. In this way it seems possible to map each types of HATI long and short orbits onto single CQSFA orbits. It is unlikely that the  directly-recolliding orbit variants are present in the SFA, as they require a collision before the laser has changed sign.

Using our new method of solving the saddle point solutions (using previous orbit types to solve the next), we are able to carefully choose initial conditions such that we can probe these orbits deriving from different energy ranges. If we start the solver in a medium energy range, $|p_f|\approx 1.6 \sqrt{\up}$ panel a), we find the standard orbits that return after a single laser cycle and scatter off the core, as shown in Fig.~\ref{fig:OrbTypes}. If we start the solver in a low energy region, $|p_f|\approx 0.3 \sqrt{\up}$ panel b), we find orbits that softly scatter after multiple laser cycles and are deflected or softly scattered by the core. However, if we start the solver in a high energy region, $|p_f|\approx 3.0 \sqrt{\up}$ panel c), we find the directly-recolliding orbits.

\section{Cutoff Comparisons}
\label{sec:cutoff}
\begin{figure}
	\includegraphics[width=\textwidth]{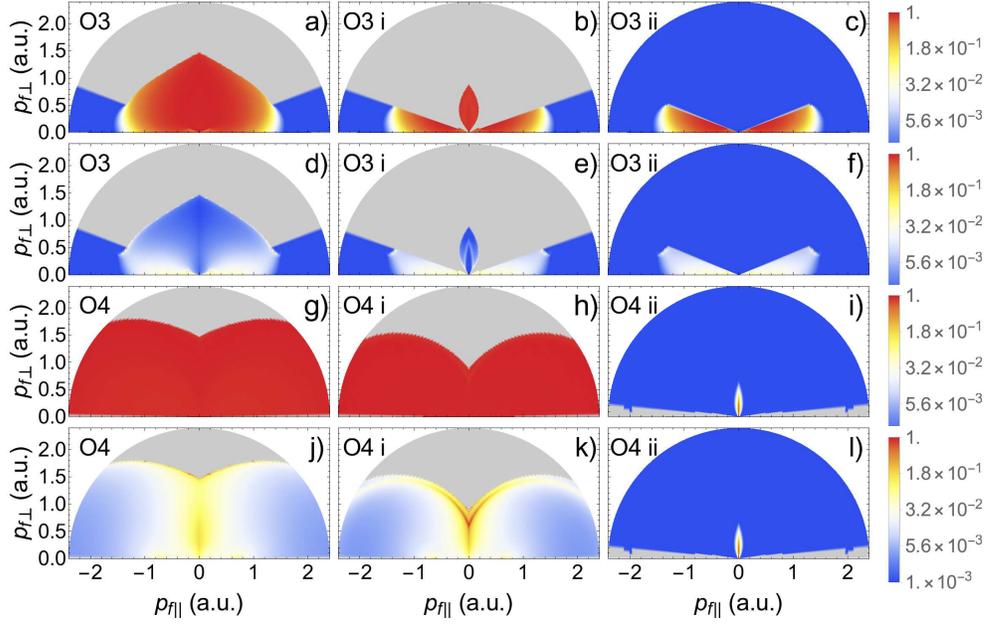}
	\caption{Single-orbit probability distributions for standard, multi-pass and directly recolliding orbits 3 and 4 (left, middle and right column, respectively). The orbit type is labelled in the top left corner. The first and second row give the distributions related to orbit 3  without and with prefactors, respectively. The third and fourth row show the distributions related to orbit 4 without and with prefactors, respectively. All distributions are normalised by their peak intensity. A logarithmic scale has been used. The same field and atomic parameters have been used as in Fig.~\ref{fig:AllOrbs}.}
	\label{fig:Regions}
\end{figure}
In this section, we have a closer look at the contributions from different types of recolliding orbits to the photoelectron momentum distributions. This includes not only their shapes but the momentum regions in which they are dominant or even present.
Fig.~\ref{fig:Regions} shows CQSFA single-orbit probability distributions computed for the three different types of orbit 3 and 4 presented in Fig.~\ref{fig:OrbTypes}. This is a good indicator of the regions in which they are important. The gray areas in the figure mark drastic topological changes in the orbits, for which other asymptotic expansions will be necessary. Within our framework, they are closely associated, but are not the sole indicator, of cutoffs. For examples of topological changes and cusps near the SFA cutoffs see, e.g.,  \cite{Faria2002}. 

Panels  a) to f) show the contributions from the three variants of orbit 3. The standard case of a single pass orbit 3 displayed in panels a) and d) has been already addressed elsewhere \cite{Maxwell2017,Maxwell2018}. These contributions occupy a large momentum range, but are restricted for large transverse momenta. The prefactor concentrates these distributions along the $p_{f\parallel}$ axis.  If orbit 3 has two passes, the corresponding probability distributions occupy a much more restricted region close to the axes, whose V shape resembles the LES reported in \cite{Hickstein2012}. Similar structures are also present for the contributions of single-pass or direct recolliding orbit 3 variants (see panels a) and c), respectively). The gray regions in panels a) and b) that start at the angle $0.1\pi$ corresponds to the transition from a soft recollision to a hard scattering event. In fact, if this angle is transformed such that $p_{0\perp}>0$ and $z_0>0$, we find $\theta_f\approx1.1 \pi$, i.e. the same angle at which orbit 3 begins to qualitatively change to become more like HATI (see discussion in the previous section). In panels b) and c), this angle marks a sharp cutoff. One should note, however, that despite the probability drop there is no topological change in the directly recolliding type 3 orbits, whose contributions are displayed in panels c) and f). Physically, this may be understood as hard scattering will always take place in this case, regardless of the photoelectron energy.
%
%

In panels g) - l)  we present the single-orbit distribution for orbit 4. The probability distribution associated with this orbit has not been studied before. If the prefactors are excluded (panel g)), the orbit has a very large flat probability distribution that extends beyond the parameter range of interest. Once the prefactors have been added (panel j)), it is restricted to mainly around the $p_{\perp}$ axis, with some small spots on the $p_{\parallel}$ axis. There is a sharp V-shaped cutoff and cusp that is similar to those found for backscattered HATI orbits in \cite{Becker2015}. 
The directly rescattering orbit 4 probability plots show only a small tulip shaped distribution with and without prefactors  (panel i) and l)). This resembles the tulip shape in panel b), for the probability plots associated with the multi-pass orbit 3. The tulip shapes that arise for both orbits 3 and 4 look similar to the cutoffs found for backscattered and forward scattered orbits in \cite{Becker2015}.

\begin{figure}
	\includegraphics[width=\textwidth]{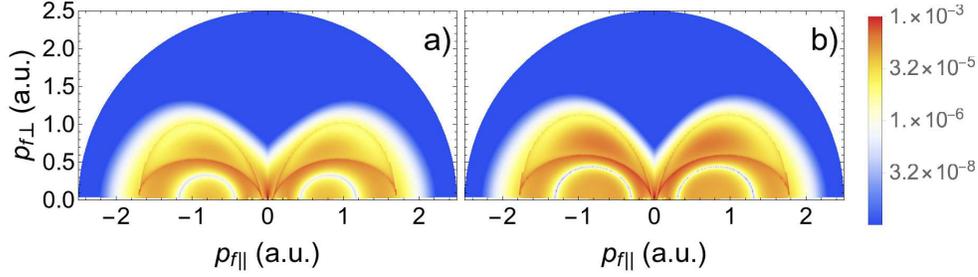}
	\caption{Probability distributions computed using HATI backscattered orbits. Panel a) and b) correspond to the shortest and second shortest possible backscattered orbit pairs starting from first half cycle. All distributions are normalised by their peak intensity. A logarithmic scale has been used. The same field and atomic parameters have been used as in Fig.~\ref{fig:AllOrbs}.}
	\label{fig:RescatteredCutoff}
\end{figure}


\begin{figure}
	\includegraphics[width=\textwidth]{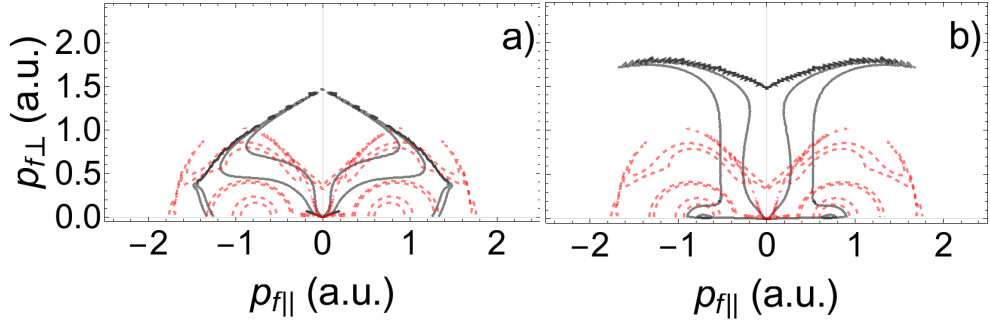}
	\caption{Comparison of CQSFA (black) and HATI (red dashed) contours, including the prefactors, using the HATI contours from Fig.~\ref{fig:RescatteredCutoff} a). Panels a) and b) relate to the standard orbits 3 and 4, respectively. The same field and atomic parameters have been used as in Fig.~\ref{fig:AllOrbs}.}
	\label{fig:ContourCompare}
\end{figure}
In order to compare the regions outlined by the single-orbit distributions in Fig.~\ref{fig:Regions} with those from the HATI case, we combine pairs of orbits using the uniform approximation in  \cite{Faria2002}. Since, in \cite{Becker2015}, it is emphasized that the Coulomb scattering cross section plays a very important role in this regime, we include the ionization and rescattering prefactors associated with the Coulomb potential. 
 
In Figs.~\ref{fig:RescatteredCutoff} a) and b) we plot the probability densities obtained with the first and second shortest HATI pair of backscattered orbits that are first sent in the opposite direction to the detector, respectively. This means that the resulting probability distributions will share features that are common to both orbits 3 and 4. Striking examples are the V-shaped structure near the $p_{f\perp}$ axis and cusps in the low-energy region.
These distributions are compared in more detail in Fig.~\ref{fig:ContourCompare}, where the contours related to the HATI distributions in Fig.~\ref{fig:RescatteredCutoff} and the CQSFA orbit 3 and 4 distributions are plotted. Fig.~\ref{fig:ContourCompare}a) shows that the shapes determined by the backscattered HATI orbits and orbit 3 from the CQSFA are not so different. They both have off-centre distorted ellipses/side-lobes along the parallel momentum axis. There is also a V-shaped structure where these distorted ellipses meet in both models. Despite the fact that orbit 3 is forward scattered and the HATI orbits are backscattered, one should note that both types of orbits are first displaced in the opposite direction  to the detector. A striking feature is that the cusps near the origin are exactly at the same place. In HATI, the cusps are due to Stokes transitions, and if the Coulomb potential is taken into consideration, this corresponds to the region in which LES have been identified.
This strongly suggests that the appropriate asymptotic expansion will change in this region for the CQSFA. 
In panel b), one can clearly see that, despite being backscattered, orbit 4 leads to very different distributions than those obtained for HATI backscattered orbits. However, there is some similarity on the $p_{f\parallel}$ axis in the form of two off-centre spots. This all indicates that the initial direction followed by the electron is more important than the type of scattering it undergoes.  

\section{Interferences in the CQSFA}
\label{sec:interf}
\begin{figure}
	\includegraphics[width=\textwidth]{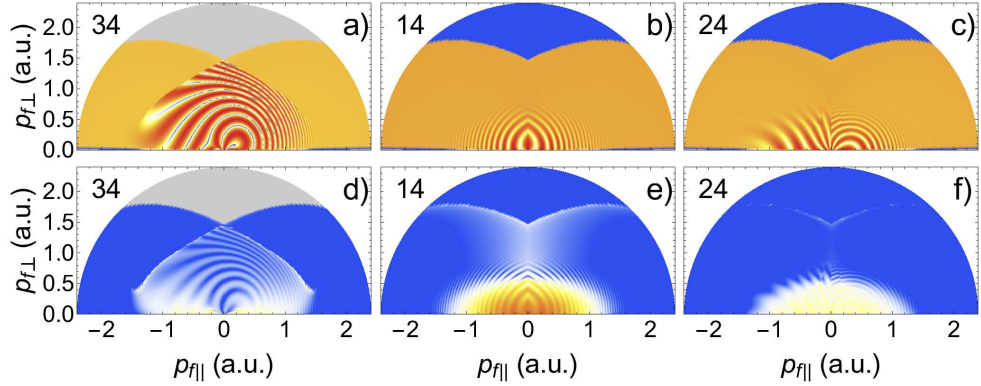}
	\caption{Photoelectron angular distributions computed using pairwise combinations of the first three types of orbits with orbit 4 to produce interference patterns. The orbit combinations are labelled in the top left corner. The upper and the lower panels neglect and incorporate prefactors, respectively. The left, middle and right column employ orbits 3 and 4, 1 and 4, and 2 and 4, respectively.  All distributions are normalised by their peak intensity. A logarithmic scale has been used. The same field and atomic parameters have been used as in Fig.~\ref{fig:AllOrbs}.}
	\label{fig:Pair}
\end{figure}

In Fig.~\ref{fig:Pair} we plot interference patterns from combining each orbit with orbit 4, which have been neglected until now. Panel a) shows interference between orbits 3 and 4, which gives rise to a spiral-like pattern. Including the prefactors, as shown in panel d), causes the signal from orbit 3 to be mainly located on the $p_{f\parallel}$ axis and that from orbit 4 to be mostly along the $p_{f\perp}$ axis. Since there is not much overlap between these regions, the interference fringes are relatively faint.

The fringes' faintness, along with the fact they could be confused with intercycle interference rings, may explain why they have been overlooked in experiments or other theoretical computations. In panel b) we see the interference pattern that arises from orbits 1 and 4. Once prefactors have been included (panel e)), the V-shaped structure is very distinctive. The interference fringes are truncated circles, which also may be confused with  ATI rings. Interference between orbits 2 and 4 (panels c) and f)) also leads to a faint V-shaped structure. The fringes for the mid-energy orbit 4 trajectories plot are fan-like on the left side and resemble off centre circles on the right. 

\begin{figure}
	\includegraphics[width=\textwidth]{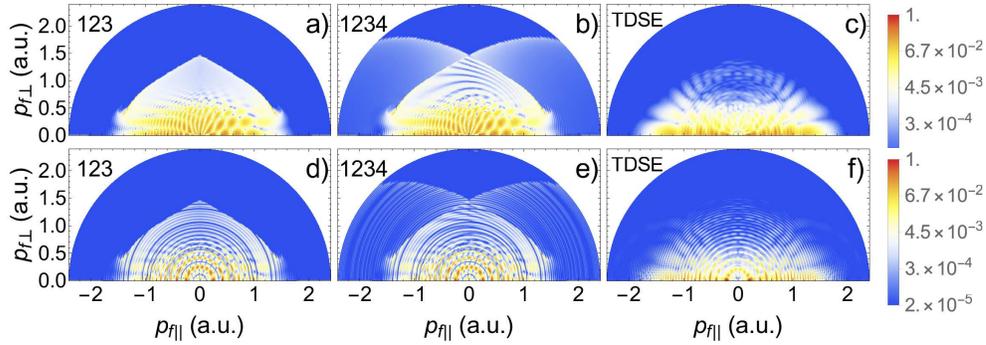}
	\caption{Full CQSFA photoelectron angular distributions calculated excluding and including orbit 4 (left and middle columns, respectively), compared with the TDSE solution (right column).  The distributions in the top and bottom row have been calculated over a single and four laser cycles, respectively. The orbits included in each distribution are marked in the top left, and the prefactors have been included in all cases. The freely available software Qprop \cite{BauerQprop2006} was used to perform the TDSE calculation. All distributions are normalised by their peak intensity. A logarithmic scale has been used. The same field and atomic parameters have been used as in Fig.~\ref{fig:AllOrbs}.}
	\label{fig:AllOrbAllCycle}
\end{figure}
If we compute the photoelectron angular distributions combining all orbits and compare them to those obtained with the coherent superposition of the first three, we can see what effect orbit 4 has. This has been done in Fig.~\ref{fig:AllOrbAllCycle} for one and four laser cycles, in which the CQSFA is also compared with the ab-initio solution of the time-dependent Schr\"odinger equation (TDSE) \cite{BauerQprop2006}. Comparing panels a) and b) we can see that adding orbit 4 does little to change the central fringes that are dominated by the fan- and spider-like structures. These structures are well known in the literature and have been discussed in previous publications \cite{Lai2017,Maxwell2017,Maxwell2018}. They are mainly due to the interference of orbits 1 and 2, and 2 and 3, respectively. However, above this region there are clear spiral fringes, which are also visible near the $p_{f\perp}$ axis for the ab-initio solutions (see panel c)). Additionally, the V-shaped structure is very visible in the high-energy region near the $p_{f\perp}$ axis for the CQSFA.

If four laser cycles are taken into account, as shown in panels d) and e), again there is little change to the main fringes. However, the coherent superposition of the spiral-like patterns and ATI rings causes chopped up fringes that appear to be interlocking. This more closely matches solutions from the TDSE in this region, shown in panel f), where the ATI rings are not solid but exhibit some interlocking gaps. 
The V-shaped structure is not explicitly identifiable in the TDSE results. However, the inclusion of orbit 4 introduces a faint signal in the very high energy regions,and improves the agreement with the TDSE results.

\section{Conclusions}
\label{sec:concl}
We have performed an in-depth analysis of recollision in the Coulomb distorted orbits that arise in the CQSFA model \cite{Lai2015,Lai2017,Maxwell2017,Yan2010a,Yan2012}. This includes (a) understanding the various types of collision that are present in the CQSFA orbits, (b) a comparison with direct and rescattered ATI orbits within the scope of the standard SFA, (c) an extensive discussion of additional types of back- and forward scattered trajectories, and (d) their influence on holographic patterns that form in ATI photoelectron angular distributions (PADs).

We use the distance $r_c$ of closest approach of an electron to the core to determine whether a specific Coulomb-distorted trajectory is a direct, softly recolliding or hard scattered orbit. 
If this distance is smaller than the Bohr radius, the collision is hard, and if $r_c$ is between the Bohr radius and the tunnel exit, the collision is soft. The classification we use differs from that in \cite{Kastner2012,Kastner2012PRL,Kelvich2016}, where the separation of zeros in the transverse coordinate determine the collision type. However we still yield the same class of softly re-colliding orbits responsible for the LES and VLES as found in \cite{Kastner2012,Kastner2012PRL,Pisanty2016,Kelvich2016,Liu2010a}.
 We find that some CQSFA orbits have analogues in direct and rescattered ATI. Thereby, the angle $\theta_f$ associated with the final momentum determines whether the CQSFA orbit will tend to direct or rescattered ATI in the high energy limit. Low and high values of  $\theta_f$ will lead to direct and rescattered trajectories, respectively, while for angles in between we find softly colliding orbits. It is worth noticing that even for relatively low photoelectron energies the CQSFA orbits behave like rescattered ATI. They undergo hard collisions, which take place very close to the core, with similar ionisation and rescattering times. This is the reason why many low-energy structures can be explained by standard rescattered SFA orbits \cite{Becker2015}.
 
 Our analysis also shows that the classification introduced in \cite{Yan2010a} for Coulomb-corrected methods, which singles out four types of orbits, is an over-simplification. We have in fact identified several topologically different orbits that would fit under a single type. These include multi-pass orbits that leads to cusps in the low-energy ATI region. A proper treatment of these structures in ATI photoelectron distributions, however, will require the development of novel asymptotic expansions and is beyond the scope of this work. Nonetheless, we have identified angular regions for which the topology has changed. These features are also present for orbits undergoing single hard collisions. The limitations of the CQSFA for these orbits have been pointed out in our previous work \cite{Maxwell2017}.

 We go further than previous studies, in that we include the backscattered electron trajectory, which, under the classification in \cite{Yan2010a} is known as orbit 4, and obtain formerly overlooked interference patterns. In previous publications the signal of orbit 4 was considered too low for it to have any effect on the photoelectron momentum distribution \cite{Lai2015,Lai2017,Maxwell2017}. However, we have shown that, although orbit 4 has little effect on medium energy ranges, beyond this the interference between orbits 3 and 4 produces a spiral-like pattern. This pattern interplays with the inter-cycle ATI-rings to make interlocking fringes, which are visible in the TDSE solution. There is also a V-shaped cusp in the very highest energy region spanned by orbit 4, that cannot be seen with the same prominence in the TDSE results. However, a similar V-shaped structure was found in \cite{Becker2015}, even though, therein, this feature occurs in a different parameter range. 
 
 One should also note that, while in the rescattered SFA the orbits occur in pairs, which are defined around a field crossing as long and short and almost merge at local energy maxima \cite{Faria2002}, the presence of the Coulomb potential disrupts this pattern and introduces additional complexities that are not fully understood. We have however verified that the orbits in the CQSFA become degenerate along the axes, and exploited this property to find solutions for the corresponding saddle point equations. This is what has enabled a solution of orbit 4 and also made it easier to find multi-pass orbits. It should be noted that the degeneracies at the boundaries will invalidate the use of the standard saddle point approximation, which is only applicable for well separated saddles. However, the size of the region for which the approximation is no longer valid is possibly quite small, as these boundaries can be asymptotically approached without issue. 
 Despite the above-mentioned differences, we have identified some similarities between the PADs computed with the standard rescattered SFA and the CQSFA. The HATI cutoffs, as found in \cite{Becker2015}, for orbits 3 and 4, lie within similar ranges.  This is expected as both types of orbits share key features that will influence the momentum regions identified by both types of distributions. This possibly the reason why a Coulomb-distorted scattering prefactor such as that employed in \cite{Becker2015} reproduces key features encountered in experiments. 

\noindent \textbf{Acknowledgements:} This work was supported by the UK EPSRC (grant EP/J019240/1). We would like to thank W. Becker for calling Ref. \cite{Becker2015} on low-energy rescattered ATI to our attention.

\end{document}